# Effects of compocasting process parameters on microstructural characteristics and tensile properties of A356−SiC$_p$ composites

Hamed Khosravi, Hamed Bakhshi, Erfan Salahinejad

Faculty of Materials Science and Engineering, K.N. Toosi University of Technology, Tehran, Iran

**Corresponding author:** Hamed Khosravi; E-mail: hkhosravi@mail.kntu.ac.ir

## Abstract

The effects of compocasting process parameters on some structural and tensile characteristics of the A356−10% SiC$_p$ (volume fraction) composites were studied. Semisolid stirring was carried out at temperatures of 590, 600 and 610 °C with stirring speeds of 200, 400 and 600 r/min for 10, 20 and 30 min. The distribution of the SiC particles within the matrix, porosity content and tensile properties of the obtained samples were examined. The structural evaluations show that by increasing the stirring time and decreasing the stirring temperature, the uniformity in the particle distribution is improved; however, by increasing the stirring speed the homogeneity firstly increases and then declines. It is also found that by increasing all of the processing parameters, the porosity content is enhanced. From the tensile characteristics viewpoint, the optimum values of the speed, temperature and time are found to be 400 r/min, 590 °C and 30 min, respectively. The contribution of the reinforcement distribution uniformity prevails over that of the porosity level to the tensile properties.

## 1. Introduction

Metal-matrix composites (MMCs) are widely used in various industries, such as aerospace, aircraft, automotive, agriculture, mining and manufacturing. The impetus behind the development of MMCs is the ability to obtain a desired combination of properties which are not obtainable in monolithic materials. The addition of high strength and high-modulus particles to a ductile metal matrix produces a material whose mechanical properties are intermediate between the matrix alloy and the ceramic reinforcement. Recently, MMCs have received substantial attention because of their improved strength, high elastic modulus, low thermal expansion coefficient and increased wear resistance over conventional based alloys. For these materials, silicon carbide (SiC) particles have become the main type of reinforcement used due to their good compatibility with aluminum alloys together with their low cost as well as excellent thermal, physical and mechanical properties. However, the increase in strength and stiffness occurs at the expense of toughness [1−7].

MMCs are produced via different routes, mainly casting and powder metallurgy techniques. The fabrication of particulate reinforced MMCs by casting methods is commercially practiced over the past decades, due to its potential in terms of production capacity and cost efficiency. Among casting techniques, stir casting is the most frequently used route to produce particulate MMCs. However, it is associated with some inherent problems arising mainly from both the apparent non-wettability of ceramic reinforcing particles by liquid aluminum alloys and the density differences between the two phases [8,9]. In order to overcome some of these drawbacks that result in the non-uniform distribution of the reinforcement within the matrix alloy, extensive interfacial reactions and formation of brittle phases at the particle/matrix interface as well as a high level of porosity, new semi-solid processing techniques have been considered for manufacturing these MMCs [10−12].

Compocasting is a semi-solid processing route in which ceramic reinforcing particulates are added to a semi-solid matrix alloy via mechanical stirring. This technique has two different variations, namely semisolid−semisolid (SS) and semisolid−liquid (SL). In both of the variations, the matrix alloy during the mixing step is in the semi-solid state, while the matrix during the casting step is partially liquid or fully liquid for the SS and SL routes, respectively. The major drawbacks of the SS process are porosity and processing difficulties, due to high viscosity, to produce defect-free castings with a uniform particle distribution [13,14].

One of the main challenges associated with the cast MMCs is to achieve a homogeneous distribution of reinforcement within the matrix alloy. In order to achieve the optimum properties of the MMCs, the distribution of the reinforcement particles in the matrix alloy should be uniform and the porosity levels need to be minimized. A non-homogeneous particle distribution often arises as a result of agglomeration, settling and segregation of ceramic particles during the processing of these composite materials. For obtaining a homogeneous distribution of reinforcing particles in the cast particulate MMCs, several factors such as the good wettability of the particles with the molten alloy, proper mixing, reinforcement size, reinforcement content, mold temperature and solidification rate should be considered [12,13,15−18].

A detailed study on the effects of the compocasting technique parameters on the reinforcement particle distribution is very critical to obtain the enhanced mechanical properties. Thus, in this work, A356−SiC$_p$ composites containing 10% SiC (volume fraction) particles of 20 µm in size were fabricated using the compocasting technique with different stirring temperatures, stirring durations and stirring speeds. The purpose was to study the effects of these factors on the SiC particle distribution and the porosity content of the achieved samples. Finally, the correlation between the microstructural characteristics and tensile properties was also discussed.

## 2. Experimental

Al-A356 with the nominal chemical composition listed in Table 1 was used as the matrix alloy. A356 aluminum alloy is a hypoeutectic Al−Si alloy and its relatively broad semisolid interval (32 °C) makes it suitable for semisolid processing. The liquidus and solidus temperatures of this alloy are 615 and 583 °C, respectively. A356 possesses improved mechanical properties, as compared with 356, because of the lower iron content.

**Table 1.** Chemical composition of A356 alloy (mass fraction, %)

| Si | Mg | Mn | Zn | Cu | Fe | Al |
|---|---|---|---|---|---|---|
| 6.93 | 0.38 | 0.23 | 0.26 | 0.25 | 0.11 | Bal. |

SiC particles with an average size of 20 µm were used as the reinforcing particles. The SEM image of the SiC powder is shown in Fig. 1.

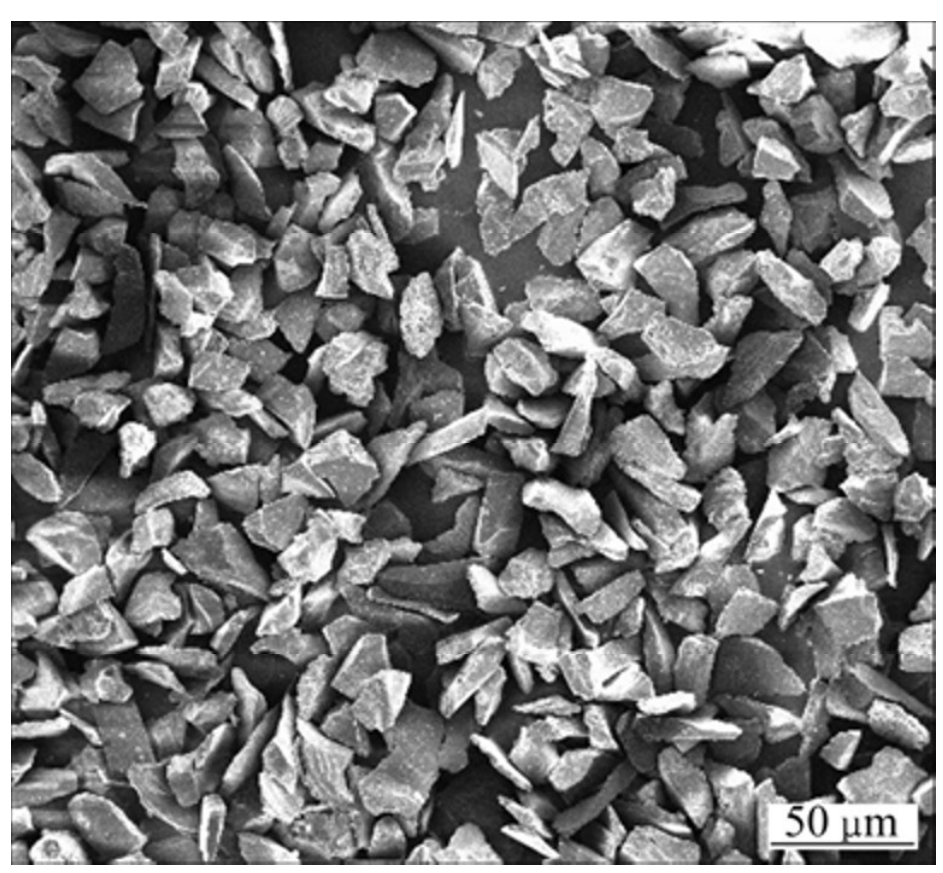

**Fig. 1.** SEM image of SiC particles

The SiC particles were artificially oxidized in air at 1000 °C for 120 min to form a layer of $SiO_2$ on them and to improve their wettability with molten aluminum. This treatment helps the incorporation of the particles while reducing undesired interfacial reactions [13].

The composites were processed by the compocasting method. At the first stage, 1 kg of A356 aluminum alloy was put in a graphite crucible and melted at 750 °C by an electric resistance furnace. Two calibrated thermocouples were inserted into the melt and the furnace to measure their temperatures. The SiC particles were preheated at 600 °C in a stainless steel crucible. Given density values for Al and SiC (2.7 and 3.2 g/cm$^3$), the crucible charge was determined to obtain A356−10% SiC composite samples. Semi-solid stirring was carried out by a graphite impeller at temperatures of 590, 600 and 610 °C for different intervals of 10, 20 and 30 min. Three stirring speeds of 200, 400, 600 r/min were also utilized. In the last stage, the slurry was heated up to 660 °C and again stirred at this temperature for 8 min. Casting was done in a cylindrical-shaped steel mold with 40 mm in internal diameter and 30 mm in height, and preheated at 400 °C.

The samples for metallographic investigations were cut from the geometrical center of the produced composite ingots. These samples were subjected to standard metallographic procedures and examined via an Olympus-BX60M light microscope. The distribution of the SiC particles within the matrix alloy was characterized by calculating the distribution factor (DF, *F*d) defined by Eq. (1) [19]:

$$DF = \frac{S.D.}{A_f} \tag{1}$$

where $A_f$ is the mean value of the area fraction of the SiC particles measured on 100 fields of a sample and S.D. is its standard deviation.

The density ($\rho$) of the produced composites was measured using Archimedes' principle. The theoretical density ($\rho_0$) of the composites was calculated from the rule-of-mixtures. The volume percentage of porosity ($\eta$) was calculated using Eq. (2).

$$Porosity\% = (1 - \frac{\rho}{\rho_0}) \times 100 \tag{2}$$

Tensile tests were conducted in a computerized testing machine (Zwick/Roell Z100) at room temperature to obtain yield strength (YS), ultimate tensile strength (UTS) and elongation (El) at a strain rate of 0.5 mm/min. Three tensile specimens with dimensions of 20 mm × 5 mm × 2.0 mm were tested for each sample.

## 3. Results and discussion

### *3.1. Effect of compocasting process parameters on microstructural characteristics*

Figure 2 demonstrates the optical micrographs of the A356−10% SiC$_p$ composites fabricated with different compocasting process parameters, including stirring temperature, stirring time and stirring speed. The quantitative assessment of the particle distribution within the composite samples was performed by considering the DF, as mentioned earlier. The smaller value of DF is indicative of the more uniform distribution of the SiC particles in the matrix [13,19]. The effect of the compocasting parameters on the uniformity of the particles distribution within the matrix is given in Fig. 3. It can be seen that the decreased stirring temperature results in a more homogeneous distribution of these particles within the matrix, as indicated by smaller DF values. The comparison of Figs. 2(d), (e) and (f) reveals that by increasing the semisolid stirring temperature, a less homogeneous distribution of the SiC particles is obtained in the matrix alloy. Decreasing the stirring temperature from 610 to 590 °C (at the fixed stirring speed and stirring time of 400 r/min and 20 min, respectively) leads to a decrease in the DF value by 45%, which is attributed to the increased viscosity of the semisolid slurry. According to the equilibrium binary Al−Si diagram, A356 aluminum alloy solidifies at a broad temperature interval (32 °C) between 583 to 615 °C. Figure 4 demonstrates that the Al−A356 alloy consists of 45%, 35% and 18% solid fractions in semisolid slurry at 590, 600 and 610 °C, respectively. This shows that the viscosity of the alloy increases with reducing the semisolid temperature. The restricted movement of the particles within the slurry during semisolid stirring as a consequence of the increased effective viscosity prevents the SiC particles from settling; consequently, a more uniform particle distribution is obtained. The presence of a solid phase in the semisolid slurry can also help the breakdown of the SiC clusters during stirring.

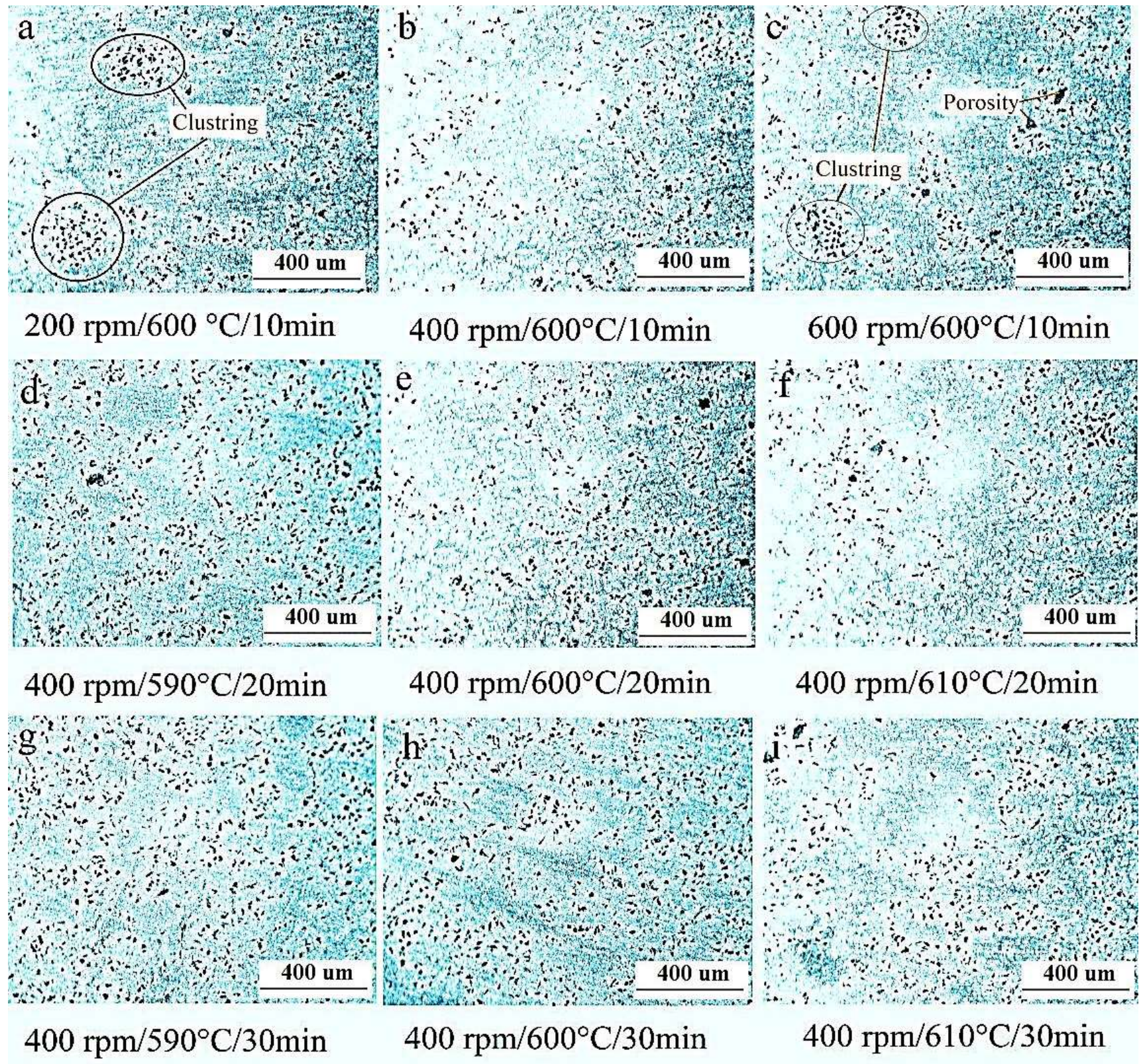

**Fig. 2.** Optical micrographs of A356−10%SiC$_p$ composites fabricated under different compocasting process conditions: (a) 200 r/min, 600 °C, 10 min; (b) 400 r/min, 600 °C, 10 min; (c) 600 r/min, 600 °C, 10 min; (d) 400 r/min, 590 °C, 20 min; (e) 400 r/min, 600 °C, 20 min; (f) 400 r/min, 610 °C, 20 min; (g) 400 r/min, 590 °C, 30 min; (h) 400 r/min, 600 °C, 30 min; (i) 400 r/min, 610 °C, 30 min.

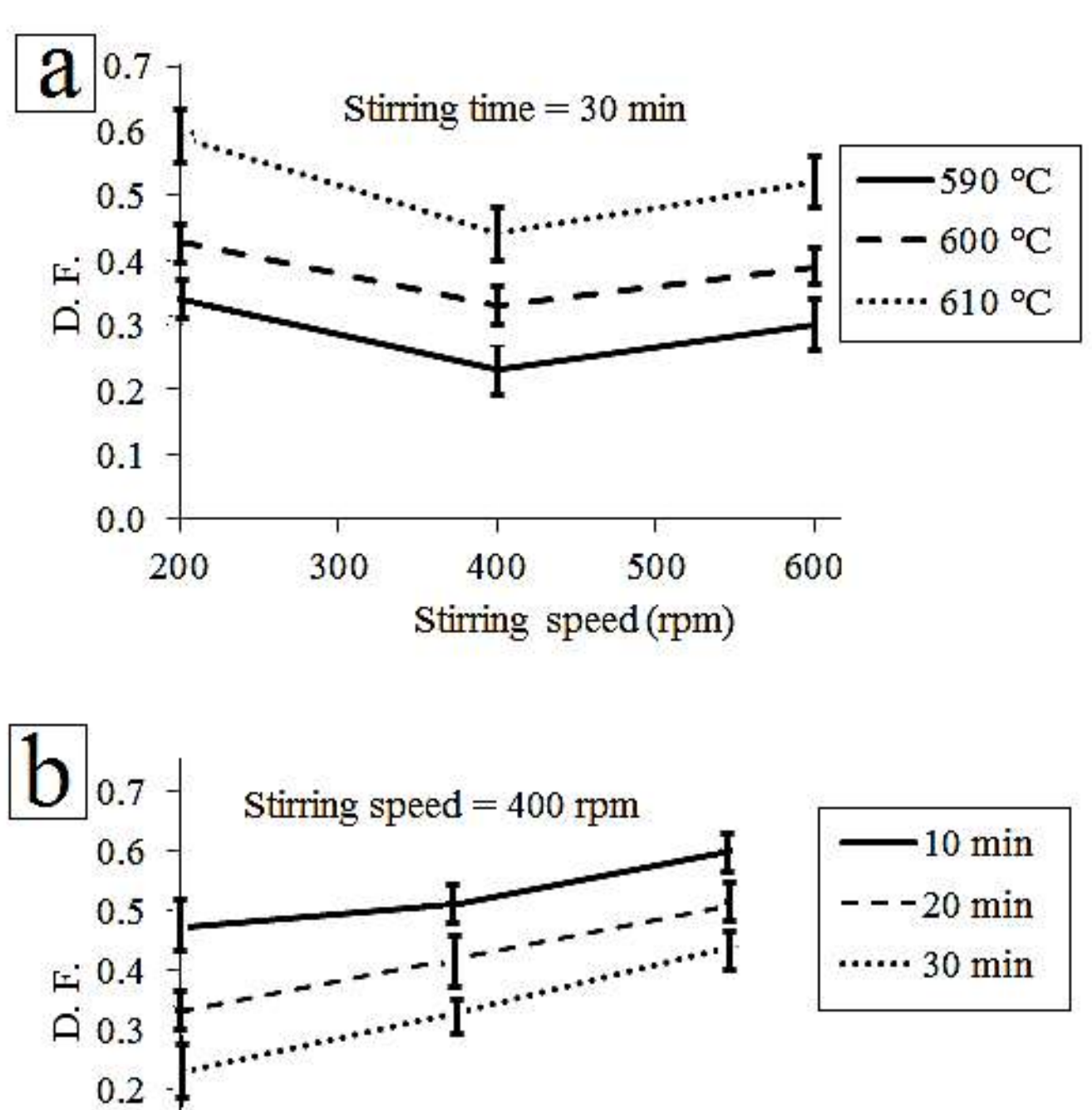

**Fig. 3.** Variation of distribution factor of SiC particles as function of compocasting process parameters.

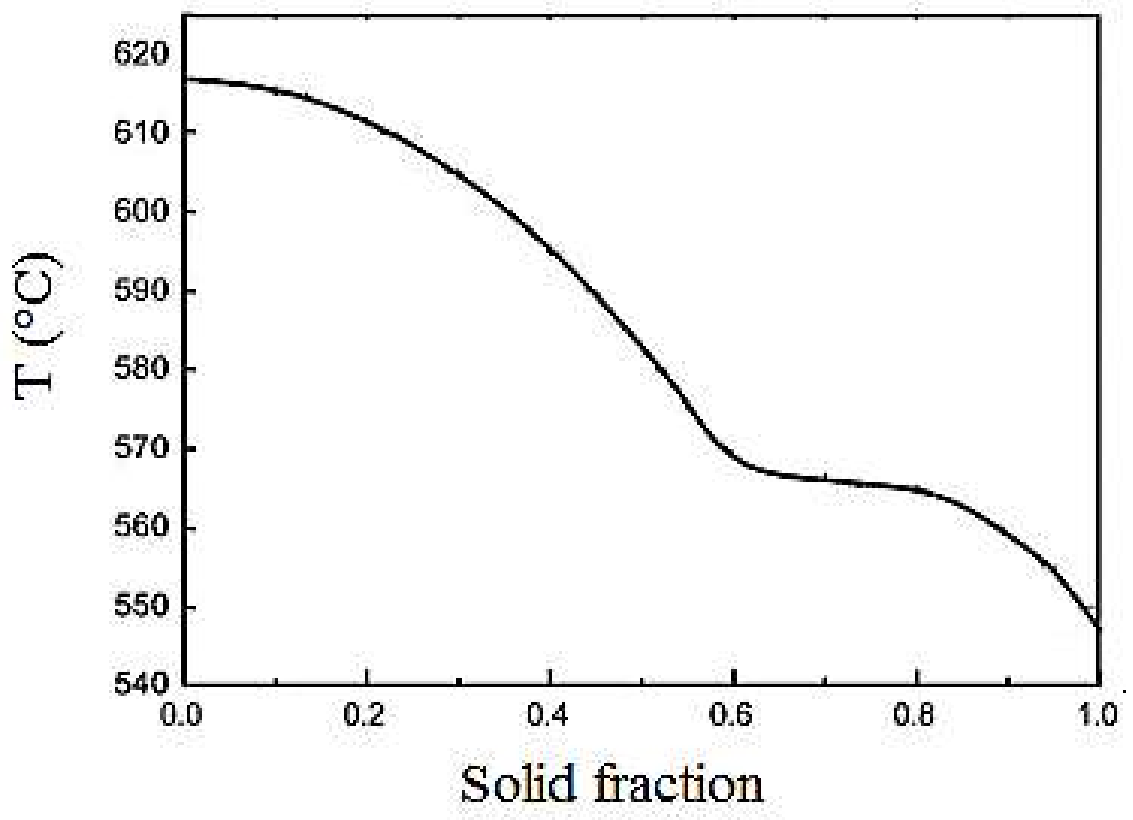

**Fig. 4.** Solid fraction of A356 alloy as function of temperature.

Figure 3(a) shows that DF of the SiC particles decrease with increasing the semisolid stirring time, representing a more homogenous SiC distribution within the matrix (Figs. 2(b), (e) and (h)). In the lower stirring time (10 min) (Fig. 2(b)), in some zones the matrix is free from the SiC particles and in other regions clustering of the SiC particles is visible. This shows that this stirring time is non-sufficient for obtaining an acceptable SiC distribution in the matrix. Compared with Figs. 2(e) and (h), the SiC particles are distributed more homogenously with increasing the semisolid stirring time from 10 to 30 min. This means that the higher stirring time results in better distribution of the particles.

The effect of the semisolid stirring speed on the degree of uniformity in the particles distribution, characterized by the DF, is given in Fig. 3(b). It can be seen that regardless of the stirring temperature and time, the DF values obtained for the stirring speed of 400 r/min are considerably smaller than those obtained for the same composites produced at the stirring speeds of 200 and 600 r/min. This shows a remarkable improvement in the uniformity of the SiC particle distribution (as reflected by the decreased DF) when the stirring speed of 400 r/min was used. At a relatively low stirring speed (i.e. 200 r/min), particle clustering is observable and in some regions, the matrix is free from the SiC particles (Fig. 2(a)). By increasing the stirring speed to 400 r/min, a better distribution of the SiC particles within the matrix alloy was obtained (Fig. 2(b)). These results are in agreement with some related studies [12,14], and can be attributed to the increase of shear forces applied by increasing the stirring speed, which can improve the uniformity of the SiC particle distribution as a result of a larger vortex within the slurry. According to Figs. 2(c) and 3(b), the higher stirring speed (600 r/min) has imposed a considerable non-uniformity in the SiC particle distribution, which can be attributed to the increased agitation severity of the slurry, resulting in clustering of the SiC particles. Therefore, it can be concluded that the modest stirring speed of 400 r/min is optimal for achieving a more homogenous distribution of the SiC particles within the matrix alloy.

The variation of porosity with the compocasting process parameters for the A356−10% SiC$_p$ composites is shown in Fig. 5. As indicated in this figure, the porosity is increased with increasing the stirring speed, time and temperature. Also, a steeper increase in porosity with increasing the stirring speed from 400 to 600 r/min is observable (Fig. 5(b)).

The porosity formation in cast metal matrix composites is influenced by a number of parameters. These include gas entrapment during stirring, air bubbles entering the slurry, water vapor on the surface of the particles, hydrogen evolution and solidification shrinkage [13,20,21]. The increased porosity of the samples with increasing the stirring time can be attributed to the increased gas absorbability and oxidation of the slurry. The decreased porosity of the samples with decreasing the stirring temperature is due to the relatively high viscosity of the composite slurry, which can reduce the air entrapment during semisolid stirring of the slurry.

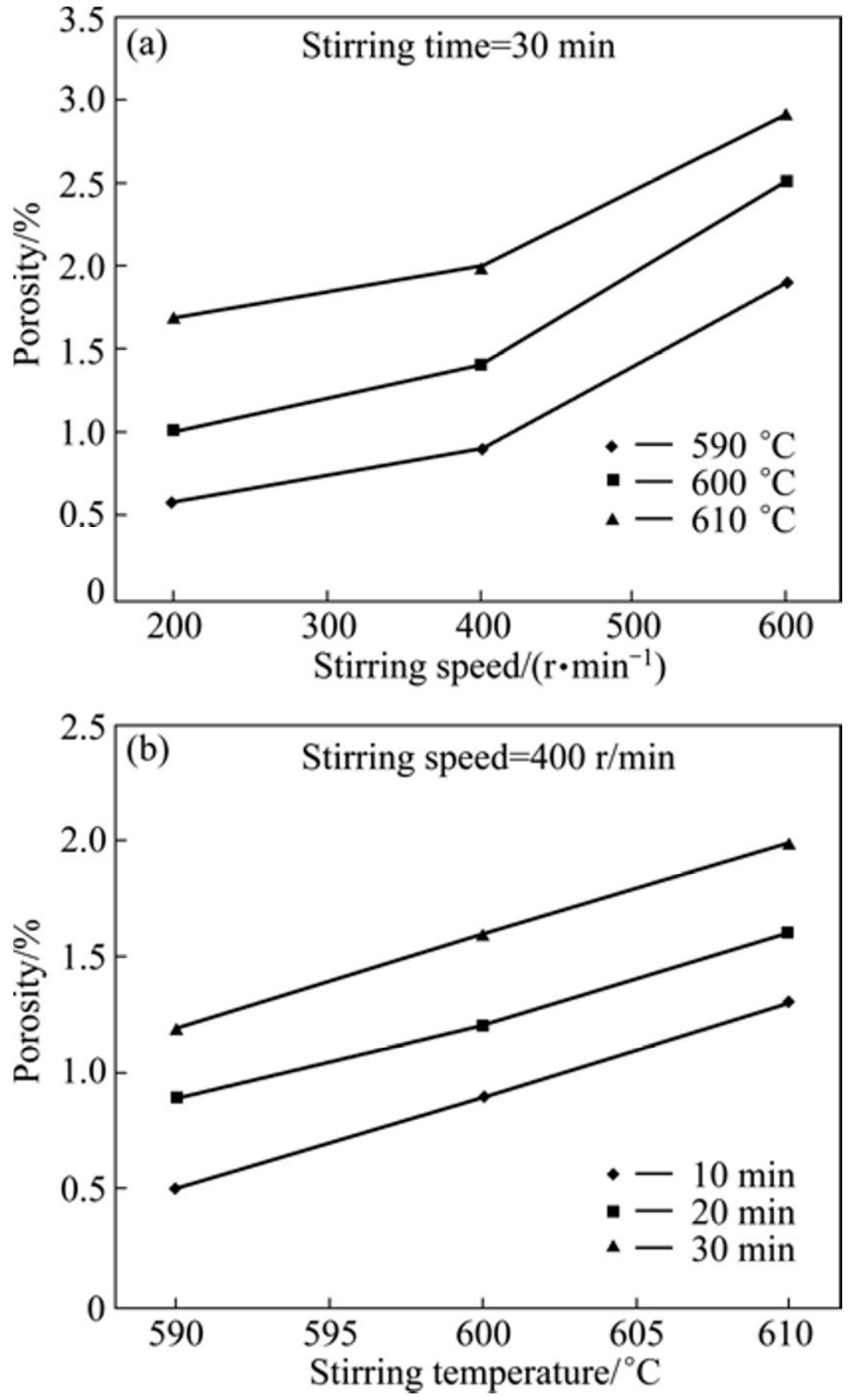

**Fig. 5.** Variation of porosity with compocasting process parameters for A356−10% SiC$_p$ composites.

As mentioned earlier, by increasing the stirring speed from 400 to 600 r/min for a fixed string time and stirring temperature the probability for the SiC particle clustering is enhanced. The air trapped in the clusters as well as the hindered liquid metal flow inside them contributes to the formation of porosity at higher stirring speeds. On the other hand, increasing the stirring speed increases the gas entrapment during semisolid stirring, thereby resulting in a higher porosity.

### *3.2. Effect of compocasting process parameters on tensile properties*

Figures 6−8 show the tensile properties of the composite samples (yield stress, ultimate tensile stress and elongation) with different compocasting process parameters. It is clear that the tensile properties of the A356−10% SiC composites increase with decreasing the increases, the tensile properties increase and approach a maximum value. Decreasing the stirring temperature from 610 °C to 590 °C (at the fixed stirring speed and stirring time of 400 r/min and 20 min, respectively) leads to the increase of yield stress, ultimate tensile stress and elongation by 19.5%, 20% and 68.8%, respectively. This can be attributed to the improved SiC distribution (Fig. 3(a)) and decreased porosity content as a result of the decreased stirring temperature (Fig. 5(a)).

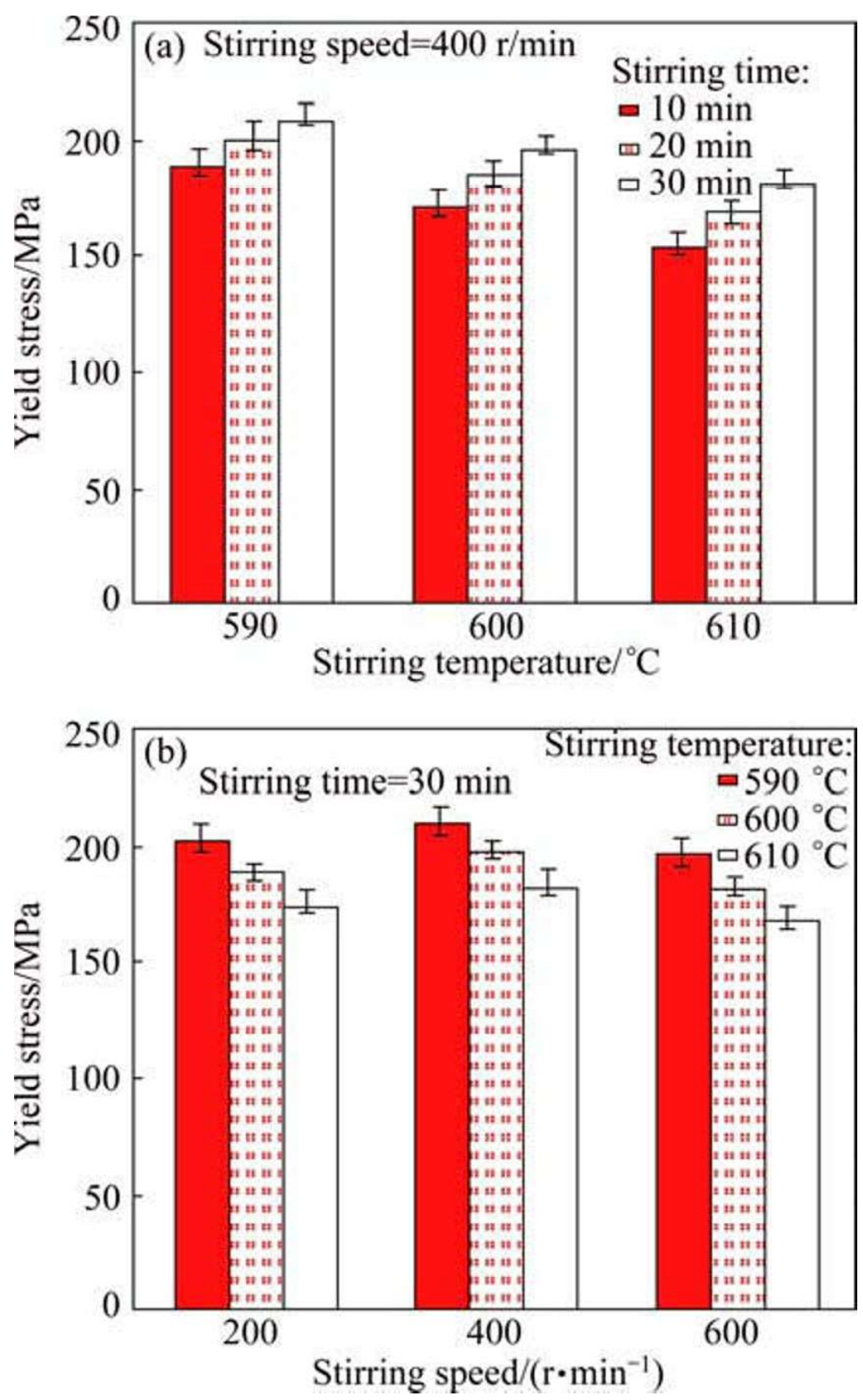

**Fig. 6.** Variation of yield stress of composites containing 10% SiC particles with different compocasting parameters.

Clustering of the reinforcement in the composite has a negative effect on the strength of the composite [7,22,23]. The clustered SiC particles are sites for damage accumulation and local particle-rich regions are the most favorable nucleation sites for crack initiation. Higher particle clustering leads to earlier fracture, due to strain localization within a particle cluster. This reduces the elongation of the composite considerably [22,24].

The improved tensile properties of the composite samples with increasing the stirring time are due to the more homogenous distribution of the SiC particles, as indicated in Fig. 3(a). Increasing the stirring time from 10 to 30 min (at the fixed stirring speed and stirring temperature and increasing the stirring temperature of 400 r/min and 590 °C, respectively) leads time. These figures also indicate that as the stirring speed to the increase of yield stress, ultimate tensile stress and elongation by 11%, 21.9% and 48.9%, respectively. With increasing the mixing time, the SiC particles are distributed more uniformly in the matrix, resulting in higher strength and ductility. As mentioned above, the porosity is increased with the increasing stirring time. Regarding the results, it can be inferred that the effect of the improvement in the SiC particle distribution dominates the adverse effect of the porosity increase on the tensile properties. It can be comparatively observed that the maximum tensile properties are obtained when the stirring speed of 400 r/min was used during the composite fabrication, regardless of the stirring time and temperature. This is attributed to more homogenous distribution of the SiC particles in this condition, as shown in Fig. 3(b). At the lower stirring speed of 200 r/min, a non-homogenous distribution of the particles in the matrix alloy was observed due to inadequate stirring. Clustering of the SiC particles as well as the high porosity level observed in the composite samples fabricated at the stirring speed of 600 r/min is believed to be reasons for the decreased tensile properties, when compared with the same composite samples produced at the stirring speed of 400 r/min. Finally, it can be concluded that the optimal compocasting process parameters to achieve the highest tensile properties in the A356−SiC$_p$ composite samples are the stirring speed of 400 r/min and the stirring temperature of 590 °C, together with 30 min of stirring.

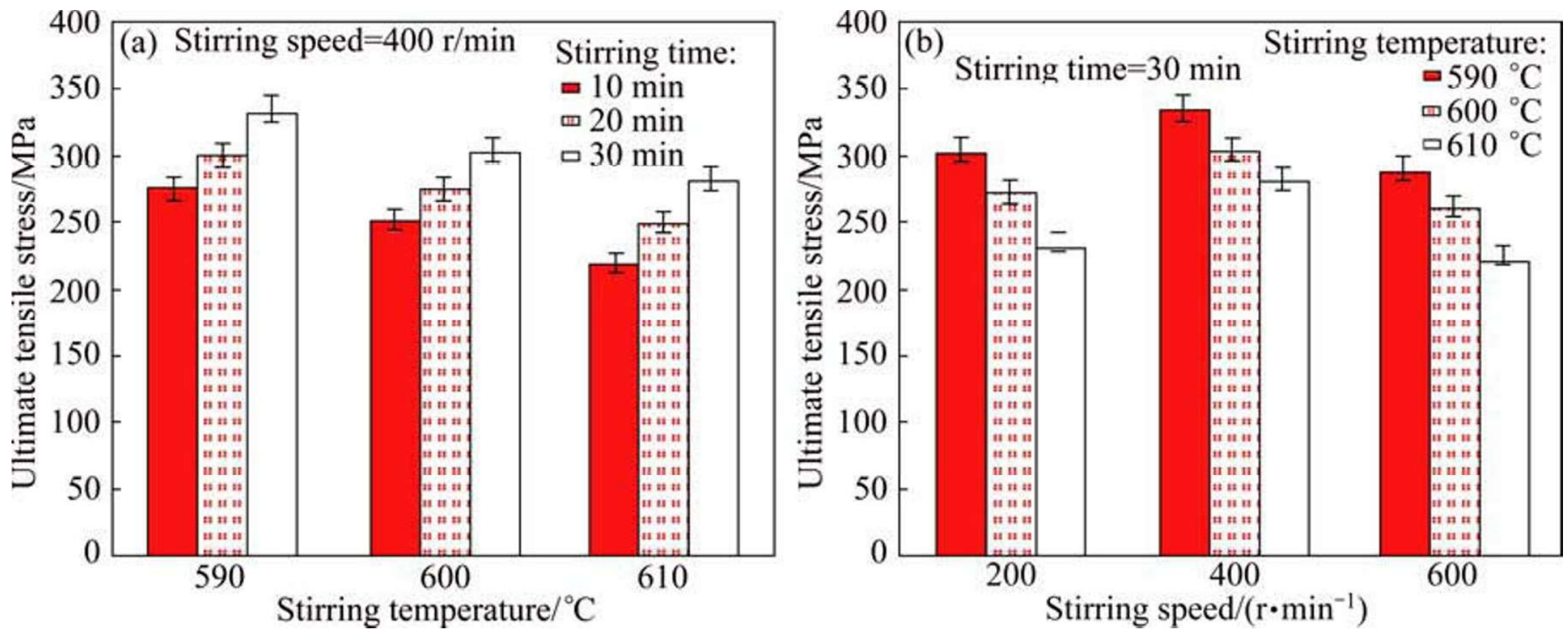

**Fig. 7.** Variation of ultimate tensile stress of composites containing 10% SiC particles with different compocasting parameters.

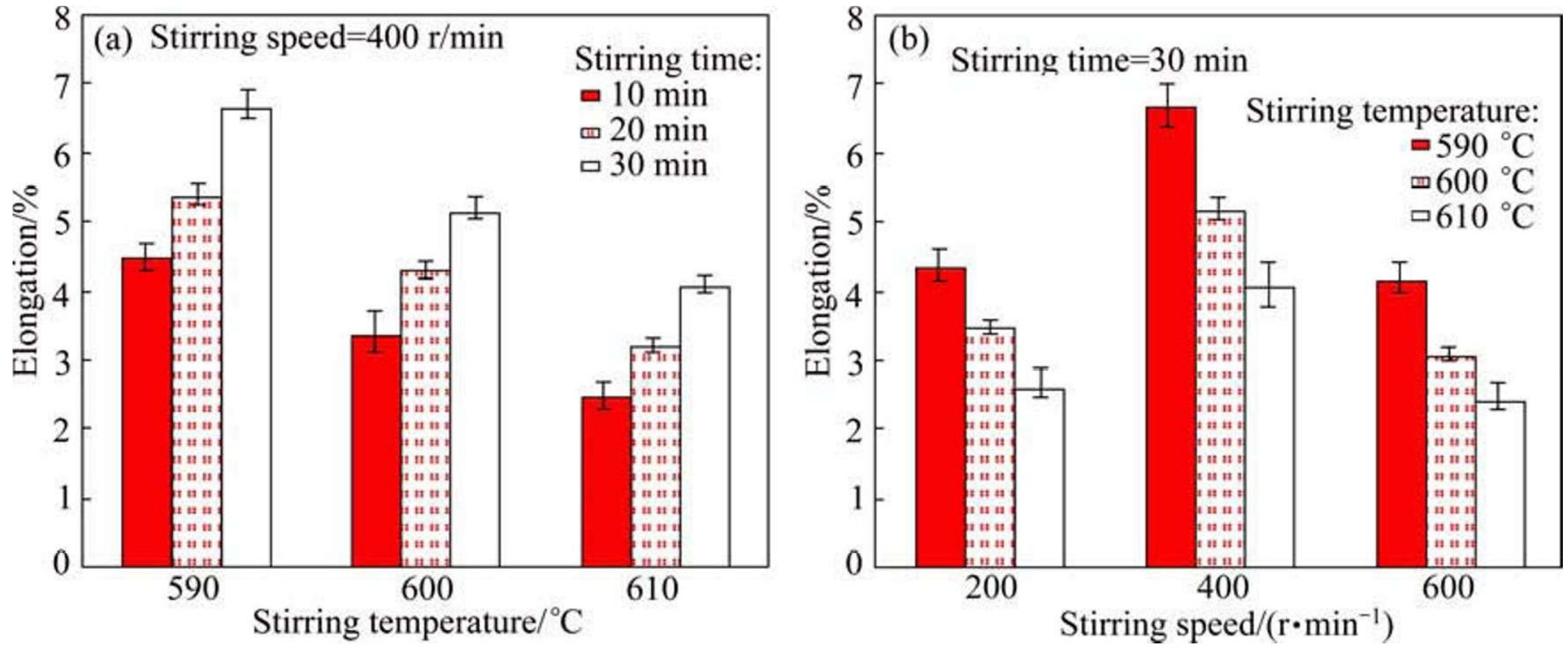

**Fig. 8.** Variation of tensile elongation of composites containing 10% SiC particles with different compocasting parameters.

## 4. Conclusions

1) By increasing the stirring time and decreasing the stirring temperature, the uniformity in the particle distribution is improved.

2) The uniformity in the particle distribution is improved by increasing the stirring speed up to an optimum level and afterward a non-uniform distribution of particles is obtained.

3) The porosity increases by increasing the stirring speed, stirring time and stirring temperature.

4) The results of the tensile tests conducted on the composite samples show that the compocasting parameters of 400 r/min, 590 °C, 30 min are the optimal values for obtaining the maximum tensile properties.